\documentclass{jpp}

\usepackage{graphicx}
\usepackage{natbib}

\ifCUPmtlplainloaded \else
  \checkfont{eurm10}
  \iffontfound
    \IfFileExists{upmath.sty}
      {\typeout{^^JFound AMS Euler Roman fonts on the system,
                   using the 'upmath' package.^^J}%
       \usepackage{upmath}}
      {\typeout{^^JFound AMS Euler Roman fonts on the system, but you
                   dont seem to have the}%
       \typeout{'upmath' package installed. JPP.cls can take advantage
                 of these fonts, if you use 'upmath' package.^^J}%
      }
  \else
  \fi
\fi

\ifCUPmtlplainloaded \else
  \checkfont{msam10}
  \iffontfound
    \IfFileExists{amssymb.sty}
      {\typeout{^^JFound AMS Symbol fonts on the system, using the
                'amssymb' package.^^J}%
       \usepackage{amssymb}%

      }{}
  \fi
\fi

\ifCUPmtlplainloaded \else
  \IfFileExists{amsbsy.sty}
    {\typeout{^^JFound the 'amsbsy' package on the system, using it.^^J}%
     \usepackage{amsbsy}}
    {\providecommand\boldsymbol[1]{\mbox{\boldmath $##1$}}}
\fi

\newsavebox{\astrutbox}
\sbox{\astrutbox}{\rule[-5pt]{0pt}{20pt}}

\newcommand\Alfven{Alfv\'en }
\newcommand\Alfvenic{Alfv\'enic }
\newcommand{\V}[1]{\mathbf{#1}} 
 
\newcommand{\zhat}{\mbox{$\hat{\mathbf{z}}$}} 
\newcommand{\xhat}{\mbox{$\hat{\mathbf{x}}$}} 
\newcommand{\yhat}{\mbox{$\hat{\mathbf{y}}$}} 
\newcommand{\khat}{\mbox{$\hat{\mathbf{k}}$}} 
\newcommand{\kphat}{\mbox{$\hat{\mathbf{k}}_\perp$}} 
\newcommand{\ea}{\mbox{$\hat{\mathbf{e}}_A$}} 
\newcommand{\eone}{\mbox{$\hat{\mathbf{e}}_{1}$}} 
\newcommand{\etwo}{\mbox{$\hat{\mathbf{e}}_{2}$}} 
\newcommand{\ep}{\mbox{$\hat{\mathbf{e}}_P$}} 

\newcommand{\secref}[1]{\S\ref{#1}}
\newcommand{\eqref}[1]{equation~(\ref{#1})}
\newcommand{\eqsref}[2]{equations~(\ref{#1})~and~(\ref{#2})}

\title[Inherently 3D Turbulence]{The Inherently Three-Dimensional Nature of Magnetized Plasma Turbulence}

\author[G. G. Howes]%
{G\ls R\ls E\ls G\ls O\ls R\ls Y\ns G.\ns H\ls O\ls W\ls E\ls S%
  \thanks{Email address for correspondence: gregory-howes@uiowa.edu}}

\affiliation{Department of Physics and Astronomy, University of
Iowa, Iowa City, IA, 52242}

\pubyear{}
\volume{}
\pagerange{}
\date{?; revised ?; accepted ?. - To be entered by editorial office}
\begin{document}

\maketitle

\begin{abstract}
It is often asserted or implicitly assumed, without justification,
that the results of two-dimensional investigations of plasma
turbulence are applicable to the three-dimensional plasma environments
of interest.  A projection method is applied to derive two scalar
equations that govern the nonlinear evolution of the \Alfvenic and
pseudo-\Alfvenic components of ideal incompressible
magnetohydrodynamic (MHD) plasma turbulence. The mathematical form of
these equations makes clear the inherently three-dimensional nature of
plasma turbulence, enabling an analysis of the nonlinear properties of
two-dimensional limits often used to study plasma turbulence. In the
anisotropic limit, $k_\perp \gg k_\parallel$, that naturally arises in
magnetized plasma systems, the perpendicular 2D limit retains the
dominant nonlinearities that are mediated only by the \Alfvenic
fluctuations but lacks the wave physics associated with the linear
term that is necessary to capture the anisotropic cascade of turbulent
energy. In the in-plane 2D limit, the nonlinear energy transfer is
controlled instead by the pseudo-\Alfven waves, with the \Alfven waves
relegated to a passive role. In the oblique 2D limit, an unavoidable
azimuthal dependence connecting the wavevector components will likely
cause artificial azimuthal asymmetries in the resulting turbulent
dynamics. Therefore, none of these 2D limits is sufficient to capture
fully the rich three-dimensional nonlinear dynamics critical to the
evolution of plasma turbulence.
\end{abstract}


\section{Introduction}
Turbulence plays an important role in mediating the transport of
particles, momentum, and energy in a wide variety of plasma
environments. From the solar interior, through the solar corona,
throughout the interplanetary medium to the magnetospheres of the
Earth and other planets, and to the interaction of the heliosphere
with the local interstellar medium, turbulence influences the
evolution of the plasma environment.  Further afield, from accretion
disks surrounding compact objects, to the interstellar medium filling
the Galaxy, to the intracluster medium within galaxy clusters,
turbulence mediates the transport of angular momentum and energy to
impact the observed appearance and evolution of each system. Finally,
in the laboratory plasmas of the magnetic confinement fusion energy program,
plasma turbulence plays a key role in limiting the efficiency of
proposed fusion reactors.  In astrophysical plasmas, the turbulence
appears to be dominated by \Alfvenic fluctuations, and it is now
widely accepted that the turbulent cascade of energy develops
anisotropically with respect to the direction of the magnetic
field. In addition, although the large-scale turbulent motions in many
astrophysical environments are sometimes adequately described by the
fluid description of magnetohydrodynamics (MHD), at the small scales
on which the turbulence is dissipated, the dynamics is often weakly
collisional, and therefore a kinetic description of the turbulent
dynamics is necessary.

The study of turbulence in a kinetic plasma, or \emph{kinetic
turbulence} \citep{Howes:2013book}, represents a new frontier in the
study of heliospheric and astrophysical plasmas. The investigation of
kinetic turbulence represents a significant challenge both
theoretically and numerically due to the six-dimensional phase space
of the kinetic description. The computational cost of kinetic
numerical simulations of astrophysical turbulence is orders of
magnitude greater than that of a comparable fluid simulation, and it
is tempting to pursue a program of turbulence simulations with reduced
spatial dimensionality. We demonstrate here that \Alfvenic turbulence
in a magnetized plasma is, however, inherently three-dimensional, and
the applicability of results based on two-dimensional simulations to
astrophysical plasma systems remains to be established.

In this study, we explore the three-dimensional nature of turbulence
in a magnetized plasma, highlighting the physical behavior that is
eliminated in treatments with reduced dimensionality.  Although our
interest is the application to realistic astrophysical plasmas that
demand a more sophisticated description of the turbulent dynamics, we
illustrate here the inherently three-dimensional nature of plasma
turbulence using the simplest case of turbulence in an incompressible
MHD plasma.  We contend that the incompressible MHD equations
represent a minimal description of the physics underlying key aspects
of plasma turbulence---the linear wave physics and nonlinear couplings
that lead to the development of an anisotropic \Alfvenic turbulent
cascade and the generation of current sheets at small scales---that
persist in more comprehensive physical descriptions. Consequently, the
three-dimensional nature of incompressible MHD turbulence is shared by
kinetic plasma turbulence.  In fact, recent laboratory experiments
have demonstrated that, under the weakly collisional plasma conditions
relevant to astrophysical environments, the nonlinear evolution of
weak plasma turbulence is well described by the equations of
incompressible MHD \citep{Howes:2012b,Howes:2013a,Nielson:2013a,
  Howes:2013b,Drake:2013}.

First, we provide simple heuristic arguments for the three-dimensional
nature of incompressible MHD turbulence in \secref{sec:simple}. In
\secref{sec:projection}, the incompressible MHD equations are
projected onto the \Alfvenic and pseudo-\Alfvenic polarizations to
obtain a set of equations that highlights the four distinct
nonlinearities, and their geometrical constraints, that determine the
evolution of a single \Alfven and pseudo-\Alfven Fourier mode. We
evaluate the nonlinear properties of different two-dimensional limits
of these equations in \secref{sec:2dlimits}, discuss the implications
of these findings for the study of plasma turbulence in
\secref{sec:discuss}, and present our conclusions in
\secref{sec:conc}.


\section{Simple Argument for Three Dimensionality}
\label{sec:simple}

A steady-state turbulent plasma is characterized by the cascade of
energy from the large scales, at which the turbulence is driven, down
to the small scales, at which the turbulence is dissipated. The
theoretical concept of an energy cascade is most easily understood by
employing a spatial plane-wave decomposition of the turbulent
fluctuations into a Fourier series, in which the cascade of energy
flows from wave modes with small wavenumbers (corresponding to the
large-scale motions) to wave modes with large wavenumbers
(corresponding to the small-scale motions).  Mathematically, the
turbulent transfer of energy is governed by the nonlinear terms in the
system of equations. These nonlinearities determine the transfer of
energy from one Fourier mode to another.  Therefore, the mathematical
properties of the nonlinear terms provide insight into the fundamental
nature of plasma turbulence. The inherently three-dimensional nature
of magnetized plasma turbulence can be simply illustrated by examining
the properties of the linear and nonlinear terms in the incompressible
MHD equations.

The ideal incompressible MHD equations can be expressed in the symmetrized
Elsasser form \citep{Elsasser:1950},
\begin{equation}
\frac{\partial \V{z}^{\pm}}{\partial t} 
\mp \V{v}_A \cdot \nabla \V{z}^{\pm} 
=-  \V{z}^{\mp}\cdot \nabla \V{z}^{\pm} -\nabla P/\rho_0,
\label{eq:elsasserpm}
\end{equation}
\begin{equation}
\nabla\cdot  \V{z}^{\pm}=0,
\label{eq:div0}
\end{equation}
where the magnetic field is decomposed into equilibrium and
fluctuating parts $\V{B}=\V{B}_0+ \delta
\V{B} $, $\V{v}_A =\V{B}_0/\sqrt{4 \pi\rho_0}$ is the \Alfven velocity 
due to the equilibrium field $\V{B}_0=B_0 \zhat$, $P$ is total pressure (thermal
plus magnetic), $\rho_0$ is mass density, and $\V{z}^{\pm}(x,y,z,t) =
\V{u} \pm \delta \V{B}/\sqrt{4 \pi \rho_0}$ are the Elsasser 
fields given by the sum and difference of the velocity fluctuation
$\V{u}$ and the magnetic field fluctuation $\delta \V{B}$ expressed in
velocity units.  Taking the divergence of \eqref{eq:elsasserpm}, the
terms on the left-hand side are zero using \eqref{eq:div0}, leaving
the following expression for the pressure,
\begin{equation}
\nabla^2 P/\rho_0 =  - \nabla \cdot \left(   \V{z}^{\mp}\cdot \nabla \V{z}^{\pm}\right)
=- \frac{\partial }{\partial x_i} \V{z}^{-}_j \frac{\partial }{\partial x_j} \V{z}^{+}_i ,
\label{eq:press}
\end{equation}
where summation over repeated indices is implied. The many 
mathematical properties of the ideal incompressible MHD equations relevant
to the study of plasma turbulence are discussed in detail in
\citet{Howes:2013a}.  Here we merely review a few  properties
relevant  to the investigation of the three-dimensional nature of plasma
turbulence. 

The second term on the left-hand side of \eqref{eq:elsasserpm} is the
\emph{linear term} that governs the lowest-order response of the
plasma to an applied perturbation.  This response is a wave-like
behavior in which the Elsasser field $\V{z}^{+}$ ($\V{z}^{-}$) travels
down (up) the equilibrium magnetic field, $\V{B}_0=B_0 \zhat$, at the
\Alfven speed, $v_A$. Henceforth, down implies the $- \zhat$
direction, and up implies the $+ \zhat$ direction. For a particular
plane-wave mode with wavevector $\V{k}=\V{k}_\perp+ k_\parallel
\zhat$, the incompressibility condition, \eqref{eq:div0}, implies that
the Elsasser fields have no variation along the direction of the
wavevector, $\V{k} \cdot \V{z}^{\pm}=0$.  Therefore, \emph{the
  Elsasser fields $\V{z}^{\pm}$ have only two nonzero components in
  the plane perpendicular to the direction of the wavevector
  $\khat$}. The two orthogonal directions that span this plane
correspond to the two types of waves in an incompressible MHD plasma: \Alfven
waves and pseudo-\Alfven waves. The direction of polarization of the
\Alfven waves is defined by $\ea \equiv \zhat \times \V{k}/|\zhat
\times \V{k}|=\zhat \times \kphat$, while the direction of
polarization of the pseudo-\Alfven waves is defined by $\ep \equiv
\V{k} \times (\zhat \times \V{k})/|\V{k} \times (\zhat \times \V{k})|
= (-k_\parallel/k) \kphat + (k_\perp/k) \zhat$. Therefore, for a
particular plane-wave mode defined by its wavevector $\V{k}$, a
natural orthonormal basis to describe the dynamics in incompressible
MHD is given by $(\hat{\V{k}},\ea,\ep)$.

On the right-hand side of \eqref{eq:elsasserpm}, the first term is the
\emph{nonlinear term} that governs the transfer of energy among plane-wave modes,
and the second term is a nonlinear pressure term that ensures
incompressibility through \eqref{eq:press}.  The mathematical form of
the nonlinear term implies that the nonlinear interaction is
nonzero only if both $\V{z}^{+} \ne 0$ and $\V{z}^{-}
\ne 0$.  Thus,  for two waves to interact nonlinearly, they must propagate
in opposite directions along the magnetic field
\citep{Iroshnikov:1963,Kraichnan:1965}. When waves are traveling
in only one direction along the magnetic field, for example when
$\V{z}^{-}=0$, an arbitrary waveform $\V{z}^{+}(x,y,z+v_At)$ is an
exact nonlinear solution of the equations, representing a finite
amplitude \Alfven or pseudo-\Alfven wavepacket traveling
nondispersively in the $-\zhat$ direction \citep{Goldreich:1995}.

The vector form of \eqref{eq:elsasserpm} readily demonstrates that the
\Alfvenic component of the turbulence in an incompressible MHD plasma
is an inherently three-dimensional phenomenon
\citep{Howes:2011a,Howes:2013a}. The linear term $\V{v}_A \cdot \nabla
\V{z}^{\pm}$ represents propagation of the \Alfven waves along the
equilibrium magnetic field and is nonzero only when the parallel
wavenumber $k_\parallel \ne 0$, requiring variation along the
field-parallel dimension. It is easily shown \citep{Howes:2013a} that
the nonlinearity arising from the interaction of counterpropagating
\Alfven waves is proportional to $\zhat \cdot (\hat{\V{k}}_\perp^-
\times\hat{\V{k}}_\perp^+)$. In order for $\hat{\V{k}}_\perp^- \times
\hat{\V{k}}_\perp^+ \ne 0$, variation in both directions perpendicular
to the magnetic field is required. This implies that both
perpendicular dimensions must be included for the nonlinear term
governing the interaction between counterpropagating \Alfven waves to
be represented properly.  Therefore, the variations in the direction
parallel to and in both directions perpendicular to the equilibrium
magnetic field must be represented to capture all of the physical
behavior at play in magnetized plasma turbulence---\emph{the dynamics
  of plasma turbulence is inherently three-dimensional, and reduction
  of the dimensionality eliminates important physical behavior}, as
will be demonstrated in detail in the remainder of this study.
Although not proven here, we speculate that the manifestly
three-dimensional nature of plasma turbulence not only applies to
incompressible MHD plasmas, but persists as a general characteristic
of the turbulence for more complex plasmas, such as compressible MHD
plasmas or kinetic plasmas.

\section{Alfv\'enic and Pseudo-Alf\'venic Projections of the Incompressible MHD Equations}
\label{sec:projection}
In this section, we explore the evolution of incompressible MHD
turbulence in terms of all possible nonlinear interactions between
counterpropagating \Alfven and pseudo-\Alfven waves. The result
illustrates clearly the inherently three-dimensional nature of
incompressible MHD turbulence.

\subsection{Evolution of a Single Fourier Mode $\V{z}^+(\V{k})$}
\label{sec:singlemode}
For simplicity in the following calculations, we choose just one of
the two symmetrized equations of ideal incompressible MHD given by
\eqref{eq:elsasserpm}, specifically analyzing the equation for the
evolution of the Elsasser field $\V{z}^{+}$,
\begin{equation}
\frac{\partial \V{z}^{+}}{\partial t} 
- v_A \frac{\partial \V{z}^{+}}{\partial z}
=-  \V{z}^{-}\cdot \nabla \V{z}^{+} -\nabla P/\rho_0,
\label{eq:elsasserp}
\end{equation}
where we have evaluated the dot product in the linear term.
We consider a triply-periodic plasma volume of size $L_\parallel \times L_\perp^2$.
The possible Fourier modes in the domain are given by 
\begin{equation}
k_{i n_i}=  2 \pi n_i/L_i \quad \quad \mbox{ for } n_i=-\infty, \ldots, -1,0,1,\ldots, \infty
\end{equation}
where $i$ signifies the spatial component $x$, $y$, or $z$, and
$L_x=L_y=L_\perp$ and $L_z=L_\parallel$.  We write the general solutions
for $\V{z}^\pm(x,y,z,t)$ as  discrete Fourier series,
\begin{equation}
\V{z}^\pm(x,y,z,t) =
 \sum^\infty_{n_x=-\infty}\sum^\infty_{n_y=-\infty}\sum^\infty_{n_z=-\infty}
\V{z}^\pm(k_{x n_x},k_{y n_y},k_{z n_z},t) e^{i \V{k} \cdot \V{r}} 
\equiv \sum_\V{k} \V{z}^\pm(\V{k},t) e^{i \V{k} \cdot \V{r}} 
\label{eq:zpmk}
\end{equation}
where $\V{k}=k_{x n_x}\xhat + k_{y n_y}\yhat + k_{z n_z}\zhat$, and the
second form is shorthand to simplify the notation.  We can also write
the total pressure $P(x,y,z,t)$ in terms of a Fourier series as well,
\begin{equation}
P(x,y,z,t) 
 = \sum_\V{k} P(\V{k},t) e^{i \V{k} \cdot \V{r}} 
\label{eq:pk}
\end{equation}
Henceforth, we suppress the explicit time dependence of the Fourier
coefficients for notational simplicity. In addition, in order for the
Elsasser fields to be real, all of the complex Fourier coefficients
must satisfy a reality condition, $\V{z}^\pm(\V{k}) =
\V{z}^{\pm*}(-\V{k})$.

Next, we substitute the Fourier series \eqsref{eq:zpmk}{eq:pk} into
\eqref{eq:elsasserp}, multiply  by $e^{-i \V{k} \cdot
\V{r}}$, and integrate the result $1/(L_\parallel L_\perp^2) \int
d^3\V{r}$ to obtain an expression for the time evolution of a single
Fourier coefficient $\V{z}^+(\V{k})$,
\begin{equation}
\frac{\partial\V{z}^+(\V{k})}{\partial t} 
- i k_\parallel v_A \V{z}^+(\V{k})
=- i\sum_{\V{k}'}\sum_{\V{k}''} \left\{ \left[\V{z}^-(\V{k}')  \cdot 
\V{k}''\right] \V{z}^+(\V{k}'') \delta^3(\V{k}'+\V{k}''-\V{k})\right\}
-\frac{i \V{k} P(\V{k})}{\rho_0}
\label{eq:zkp}
\end{equation}

\subsection{Projection onto \Alfven and Pseudo-\Alfven Polarizations}
\label{sec:proj}
In this section, we decompose a general spatial Fourier mode
$\V{z}^\pm(\V{k})$ into its \Alfvenic and pseudo-\Alfvenic components
by projection of the vector Elsasser field onto the orthonormal
basis $(\khat,\ea,\ep)$. Note that, although we refer to these
fluctuations as \Alfven and pseudo-\Alfven waves, linearization of the
equations was never performed.  This projection is applicable in the
fully nonlinear limit, and the results are therefore not limited to
small amplitude fluctuations.

Since the incompressibility condition $\nabla\cdot \V{z}^{\pm}=0$
implies that $\khat \cdot \V{z}^\pm_\V{k}=0$, the general
decomposition of the Fourier coefficient is given by $\V{z}^+(\V{k}) =
z^+_A (\V{k})\ea + z^+_P (\V{k})\ep$.  Note that the directions $\ea$
and $\ep$ are functions of the wavevector $\V{k}$, so in the following
calculations we employ a notational convention that uses $\ea'$ and
$\ep'$ to denote the \Alfvenic and pseudo-\Alfvenic directions
associated with a plane-wave mode with wavevector $\V{k}'$.

Projecting \eqref{eq:zkp} onto $\ea$  yields
a scalar equation for the evolution of the \Alfvenic component $z^+_A (\V{k})$,
\begin{equation}
\frac{\partial z^+_A (\V{k})}{\partial t} 
- i k_\parallel v_A z^+_A (\V{k})
=- i\sum_{\V{k}=\V{k}'+\V{k}''}[ \V{z}^-(\V{k}')  \cdot 
\V{k}'' ][\V{z}^+(\V{k}'')\cdot \ea]
\end{equation}
where we define a shorthand notation  to denote 
 the double sum over all possible $\V{k}'$ and $\V{k}''$ subject 
to the constraint $\V{k}=\V{k}'+\V{k}''$, 
\begin{equation}
\sum_{\V{k}=\V{k}'+\V{k}''}\equiv \sum_{\V{k}'}\sum_{\V{k}''} 
 \delta^3(\V{k}'+\V{k}''-\V{k}).
\end{equation}
Note that the pressure term drops out upon projection for both the
\Alfvenic or pseudo-\Alfvenic directions since $\khat \cdot \ea=0$ and
$\khat \cdot \ep=0$. 
Splitting  $\V{z}^-(\V{k}')$ and $\V{z}^+(\V{k}'')$ into their
\Alfvenic and pseudo-\Alfvenic components, the equation becomes
\begin{equation}
\frac{\partial z^+_A (\V{k})}{\partial t} 
- i k_\parallel \V{v}_A z^+_A (\V{k})
=- i\sum_{\V{k}=\V{k}'+\V{k}''}
\left[  z^{-}_A (\V{k}')(\ea' \cdot \V{k}'') 
+ z^{-}_P  (\V{k}')(\ep' \cdot  \V{k}'') \right] 
\left[ z^{+}_A(\V{k}'') (\ea''\cdot \ea )    
+ z^{+}_P (\V{k}'')(\ep''\cdot \ea ) \right] 
\label{eq:zpa}
\end{equation}

Performing the complementary projection of  \eqref{eq:zkp} onto $\ep$ and  following 
the same procedure to simplify the expression, we obtain 
\begin{equation}
\frac{\partial z^+_P (\V{k})}{\partial t} 
- i k_\parallel \V{v}_A z^+_P (\V{k})
=- i\sum_{\V{k}=\V{k}'+\V{k}''}
\left[  z^{-}_A (\V{k}')(\ea' \cdot \V{k}'') 
+ z^{-}_P (\V{k}')(\ep' \cdot  \V{k}'') \right] 
\left[ z^{+}_A (\V{k}'')(\ea''\cdot \ep )    
+ z^{+}_P (\V{k}'')(\ep''\cdot \ep ) \right] 
\label{eq:zpp}
\end{equation}

Note that the projection of  \eqref{eq:zkp} onto $\khat$ shows that the pressure
term cancels any component in the $\khat$ direction that is generated by the 
nonlinear term, thus maintaining incompressibility.

\subsection{Simplification of the Dot Products}
Six unique dot products appear in \eqsref{eq:zpa}{eq:zpp}: $\ea' \cdot
\V{k}''$, $\ep' \cdot \V{k}''$, $\ea''\cdot \ea $, $\ep''\cdot \ea $,
$ \ea''\cdot \ep $, and $\ep''\cdot \ep $. These dot products
represent geometrical constraints arising in each of the resulting
nonlinear terms. Using the constraint $\V{k}=\V{k}'+\V{k}''$, we
simplify these expressions as follows,
\begin{equation}
\ea' \cdot \V{k}''=k_\perp'' \zhat \cdot (\kphat' \times \kphat'')
\end{equation}
\begin{equation}
\ep' \cdot \V{k}'' = \frac{k_\perp'}{k'}  k_\parallel'' -\frac{k_\parallel'}{k'} k_\perp''(\kphat' \cdot \kphat'')
\end{equation}
\begin{equation}
\ea''\cdot \ea =\frac{k_\perp''}{k_\perp}+ \frac{k_\perp'}{k_\perp}(\kphat' \cdot \kphat'')
\end{equation}
\begin{equation}
\ep''\cdot \ea =-\frac{k_\parallel'' k_\perp'}{k'' k_\perp}\zhat \cdot (\kphat' \times \kphat'')
\end{equation}
\begin{equation}
 \ea''\cdot \ep =\frac{k_\parallel  k_\perp'}{k k_\perp}\zhat \cdot (\kphat' \times \kphat'')
\end{equation}
\begin{equation}
\ep''\cdot \ep =\frac{k_\perp k_\perp''}{k k''} + 
 \frac{k_\parallel k_\parallel'' k_\perp''}{k k'' k_\perp}
+  \frac{k_\parallel k_\parallel''k_\perp'}{k k''k_\perp}( \kphat' \cdot \kphat'')
\end{equation}
Note here that $k=|\V{k}|= |\V{k}'+ \V{k}''|$ and $k_\perp=|\V{k}_\perp|= |\V{k}_\perp' + \V{k}_\perp'' |$.
The crucial point arising from the appearance of the factors
$k_\parallel$, $(\kphat' \times \kphat'')$, or $( \kphat' \cdot
\kphat'')$ in all of these expressions is that significant simplifications
occur under reduced dimensionality.

\subsection{Final Scalar Evolution Equations for  $z^+_A (\V{k})$ and $z^+_P (\V{k})$}
Substituting the dot product relations derived in the previous section, we 
obtain the final two scalar equations that govern the evolution of a
single \Alfvenic Fourier mode $ z^+_A (\V{k})$ and of a single
pseudo-\Alfvenic Fourier mode $z^+_P(\V{k})$,
\begin{eqnarray}
\lefteqn{\frac{\partial z^+_A (\V{k})}{\partial t} 
- i k_\parallel v_A z^+_A (\V{k}) = }   \label{eq:zpa_final}\\
& - i\sum_{\V{k}=\V{k}'+\V{k}''}&
\left\{ z^{-}_A(\V{k}') z^{+}_A(\V{k}'') \left[ k_\perp'' \zhat
\cdot (\kphat' \times \kphat'') \right]
\left[ \frac{k_\perp''}{k_\perp}+ \frac{k_\perp'}{k_\perp}(\kphat' \cdot \kphat'')
\right] \right.
 \nonumber  \\
&& + z^{-}_P (\V{k}')z^{+}_A (\V{k}'') \left[   k_\parallel''\frac{k_\perp'}{k'}  - k_\perp'' \frac{k_\parallel'}{k'}(\kphat' \cdot \kphat'') \right] 
 \left[ \frac{k_\perp''}{k_\perp}+ \frac{k_\perp'}{k_\perp}(\kphat' \cdot \kphat'') \right] \nonumber  \\
&&+  z^{-}_A (\V{k}')z^{+}_P (\V{k}'')
 \left[ k_\perp'' \zhat \cdot (\kphat' \times \kphat'') \right] 
 \left[-\frac{k_\parallel'' k_\perp'}{k'' k_\perp}\zhat \cdot (\kphat' \times \kphat'')\right] \nonumber \\
&&+ \left.  z^{-}_P(\V{k}') z^{+}_P (\V{k}'') \left[   k_\parallel''\frac{k_\perp'}{k'}  - k_\perp'' \frac{k_\parallel'}{k'}(\kphat' \cdot \kphat'') \right] 
 \left[-\frac{k_\parallel'' k_\perp'}{k'' k_\perp}\zhat \cdot (\kphat' \times \kphat'') \right]\right\} \nonumber 
\end{eqnarray}
\begin{eqnarray}
\lefteqn{\frac{\partial z^+_P(\V{k}) }{\partial t} 
- i k_\parallel v_A z^+_P (\V{k}) = }  \label{eq:zpp_final}\\ 
&  - i\sum_{\V{k}=\V{k}'+\V{k}''}& \left\{
z^{-}_A  (\V{k}')z^{+}_A (\V{k}'')\left[  k_\perp'' \zhat \cdot (\kphat' \times \kphat'') \right] 
 \left[ \frac{k_\parallel  k_\perp'}{k k_\perp}\zhat \cdot (\kphat' \times \kphat'')
 \right] \right.
\nonumber 
\\
&& + z^{-}_P(\V{k}')z ^{+}_A (\V{k}'') \left[   k_\parallel''\frac{k_\perp'}{k'}  - k_\perp'' \frac{k_\parallel'}{k'}(\kphat' \cdot \kphat'') \right] 
 \left[ \frac{k_\parallel  k_\perp'}{k k_\perp}\zhat \cdot (\kphat' \times \kphat'') \right]  \nonumber \\
&& +  z^{-}_A (\V{k}')z^{+}_P (\V{k}'')\left[  k_\perp'' \zhat \cdot (\kphat' \times \kphat'') \right] 
 \left[  \frac{k_\perp k_\perp''}{k k''} + 
 \frac{k_\parallel k_\parallel'' k_\perp''}{k k'' k_\perp}
+  \frac{k_\parallel k_\parallel''k_\perp'}{k k''k_\perp}( \kphat' \cdot \kphat'')\right]  
   \nonumber \\
&& +  \left. z^{-}_P (\V{k}')z^{+}_P (\V{k}'') \left[   k_\parallel''\frac{k_\perp'}{k'}  
- k_\perp'' \frac{k_\parallel'}{k'}(\kphat' \cdot \kphat'') \right] 
\left[ \frac{k_\perp k_\perp''}{k k''} + 
 \frac{k_\parallel k_\parallel'' k_\perp''}{k k'' k_\perp}
+  \frac{k_\parallel k_\parallel''k_\perp'}{k k''k_\perp}( \kphat' \cdot \kphat'') \right]  
\right\} \nonumber 
\end{eqnarray}
The corresponding two scalar evolution equations for $z^-_A (\V{k})$
and $z^-_P (\V{k})$ may be obtained by changing the sign of the linear
term on the left-hand side and by replacing each $z^+$ with a $z^-$,
and vice versa.

Equations~(\ref{eq:zpa_final}) and (\ref{eq:zpp_final}) show
explicitly that, in each equation, there exist \emph{four distinct
nonlinearities} that dictate the interactions between
counterpropagating \Alfven and pseudo-\Alfven waves. In each equation,
the first term on the right-hand side corresponds to the nonlinear
interaction between an upward \Alfven wave and a downward
\Alfven wave, the second term to an upward pseudo-\Alfven wave and a
downward \Alfven wave, the third term to an upward \Alfven wave and a
downward pseudo-\Alfven wave, and the fourth term to an upward
pseudo-\Alfven wave and a downward pseudo-\Alfven wave.

The development of a significant wavevector anisotropy, $k_\perp \gg
k_\parallel$, for the small-scale fluctuations in magnetized plasma
turbulence is now widely accepted, based on a wide range of numerical,
experimental, and observational studies
\citep{Robinson:1971,Belcher:1971,Zweben:1979,Montgomery:1981,
  Shebalin:1983,Cho:2000,Maron:2001,Cho:2004,Cho:2009,Sahraoui:2010b,
  Narita:2011,TenBarge:2012a,Roberts:2013}. In this anisotropic limit,
a single nonlinearity dominates in each of \eqsref{eq:zpa_final}
{eq:zpp_final}. The evolution of the downward \Alfven waves governed
by \eqref{eq:zpa_final} is dominated by the first nonlinear term
corresponding to the interaction between an upward \Alfven wave and a
downward \Alfven wave. The evolution of the downward pseudo-\Alfven
waves governed by \eqref{eq:zpp_final} is dominated by the third
nonlinear term corresponding to the interaction between an upward
\Alfven wave and a downward pseudo-\Alfven wave. Therefore, in the
anisotropic limit, $k_\perp \gg k_\parallel$, the \Alfven waves are
dominantly responsible for mediating the nonlinear energy transfer
between counterpropagating wave modes. In addition, the \Alfven and
pseudo-Alfven waves do not exchange energy in this limit. The
dominance of nonlinearities which depend only on the perpendicular
gradients (neglecting the small terms involving $k_\parallel$ in the
third nonlinear term of \eqref{eq:zpp_final} above) suggests the
possibility that a two-dimensional treatment in the perpendicular
plane (the perpendicular 2D limit analyzed in \secref{sec:perp2d}) may
be sufficient to capture the turbulent dynamics. However, one of the
main points of this study is that the perpendicular 2D limit is
\emph{not} sufficient, but that the reduced, yet still
three-dimensional, description governed by the reduced MHD equations
is a preferable choice.

Equations~(\ref{eq:zpa_final}) and (\ref{eq:zpp_final}) also
demonstrate that the nonlinear energy transfer that drives the energy
cascade in incompressible MHD turbulence potentially involves all
possible triads of plane-wave modes that satisfy the wavevector
constraint $\V{k}=\V{k}'+\V{k}''$. For the general case of quadratic
nonlinearities in fluid equations, when energy is transferred to a
particular plane-wave mode with wavevector $\V{k}$, the equation for
the evolution of the mode $\V{k}$ does not uniquely determine whether
the energy originated from mode $\V{k}'$ or from mode $\V{k}''$ or
from a combination of both modes. But, in the case of the ideal
incompressible MHD equations, another property of the equations can be
used to determine uniquely the mode from which the energy was
transferred. It can be readily demonstrated that, for the ideal
incompressible MHD equations, the total energy of each of the Elsasser
fields $\V{z}^\pm$ is independently conserved
\citep{Maron:2001,Schekochihin:2009,Howes:2013a},
\begin{equation}
\frac{d}{dt} \int d^3\V{r} \  |\V{z}^\pm|^2 = 0.
\end{equation}
This property implies that the nonlinear interactions do not lead to
any exchange of energy between upward and downward waves---the energy
fluxes of the waves in each of these directions is conserved.
Therefore, since the upward waves $z^-_A (\V{k}')$ or $z^-_P (\V{k}')$
do not exchange energy with the downward waves $z^+_A (\V{k})$ or
$z^+_P (\V{k})$, any energy gained (lost) by $z^+_A (\V{k})$ or $z^+_P
(\V{k})$ must have been lost (gained) by $z^+_A (\V{k}'')$ or $z^+_P
(\V{k}'')$. In other words, the upward waves, whether \Alfven or
pseudo-\Alfven waves, merely serve to mediate the energy transfer from
one downward wave mode to another downward wave mode.  This does not
mean that the energy of an upward wave $z^-_A (\V{k}')$ or $z^-_P
(\V{k}')$ remains constant---another equation governs the evolution of
the upward wave $z^-_A (\V{k}')$ or $z^-_P (\V{k}')$ involving
nonlinear interactions mediated by downward waves.

This property can be confirmed, for example, by verifying that the
transfer of energy between two downward \Alfven waves, $z^+_A (\V{k})$
and $z^+_A (\V{k}'')$, mediated by an upward \Alfven wave $z_A^-
(\V{k}')$ in the complementary triad interactions
$\V{k}=\V{k}'+\V{k}''$ and $\V{k}''=-\V{k}'+\V{k}$ , is conservative,
given by
\begin{equation}
\frac{1}{2}\frac{\partial }{\partial t} \left( |z^+_A (\V{k})|^2 + 
|z^+_A (-\V{k})|^2 +  |z^+_A (\V{k}'')|^2 + 
|z^+_A (-\V{k}'')|^2 \right) =0.
\end{equation}
Note that the reality condition $\V{z}^\pm(\V{k}) =
\V{z}^{\pm*}(-\V{k})$ requires that all four terms above must be
included to obtain  this result.

The nonlinear triad interactions between counterpropagating \Alfven
and pseudo-\Alfven waves given explicitly in
\eqsref{eq:zpa_final}{eq:zpp_final} govern the nonlinear energy
transfer that underlies the turbulent cascade in an incompressible MHD
plasma. It is important to note that one must integrate the equations
for all of the Fourier modes self-consistently in time to determine if
any particular interaction or combination of interactions leads to a
secular transfer of energy in time.  The asymptotic analytical
solution presented in
\citet{Howes:2013a} provides an explicit example of this time integration 
that leads to a secular transfer of energy to smaller scales.
In that weakly nonlinear case, it was shown that the interaction
between two perpendicularly polarized, counterpropagating plane
\Alfven waves with equal and opposite values of $k_\parallel$ and
equal values of $k_\perp$ yields a secular transfer of energy through
a combination of two triad interactions (together comprising a
\emph{resonant four-wave interaction}): first, the two
counterpropagating \Alfven waves generate an intermediary, purely
magnetic mode with $k_\parallel=0$; second, the interaction of each of
the original \Alfven waves with that $k_\parallel=0$ intermediary
transfers energy to a third \Alfven wave with higher $k_\perp$
traveling in the same direction as the original wave. 

Although the determination of whether a particular nonlinear triad
interaction yields a net transfer of energy requires performing such a
challenging self-consistent time integration of all modes, one may
simply rule out the contribution of a particular nonlinear interaction
if the geometric factors contained in the brackets
\eqsref{eq:zpa_final} {eq:zpp_final} are zero. Indeed, it is these
geometric factors involving the wavevectors $\V{k}'$ and $\V{k}''$
that highlight the elimination, under reduced dimensionality, of many
of the four distinct nonlinearities responsible for the energy cascade
in incompressible MHD turbulence.

\section{Two-Dimensional Limits}
\label{sec:2dlimits}
Here we analyze the limits of \eqsref{eq:zpa_final}{eq:zpp_final} in
several two-dimensional limits that are often used for numerical
studies plasma turbulence. The two most widely used cases are the (i)
\emph{perpendicular 2D limit}, in which the equilibrium magnetic field
is perpendicular to the two-dimensional plane of the simulation, and
the (ii) \emph{in-plane 2D limit}, in which the equilibrium magnetic
field is contained within the two-dimensional plane of the
simulation. A few studies have employed the (iii) \emph{oblique 2D
limit}, in which the equilibrium magnetic field is inclined by a small
angle from the normal to the two-dimensional plane of the simulation.

\subsection{Perpendicular 2D Limit}
\label{sec:perp2d}

In this limit, spatial variations may occur only in the $(x,y)$ plane,
so the possible wavevectors are $\V{k}=k_x \xhat + k_y \yhat$, where
the equilibrium magnetic field is $\V{B}_0=B_0 \zhat$. With this
two-dimensional limitation of the wavevector, there is no component of
the wavevector parallel  to the equilibrium magnetic field,
$k_\parallel = \V{k} \cdot  (\V{B}_0/B_0)=0$. It therefore follows that
$\V{k}=\V{k}_\perp $, and as a result the
\Alfven wave polarization simplifies to $\ea =\zhat \times
\hat{\V{k}}_\perp$ and the pseudo-\Alfven wave polarization simplifies
to $\ep =\zhat$.  In this limit,
\eqsref{eq:zpa_final}{eq:zpp_final} undergo  significant simplifications to yield
\begin{equation}
\frac{\partial z^+_A (\V{k})}{\partial t} 
=- i\sum_{\V{k}=\V{k}'+\V{k}''} z^{-}_A (\V{k}')z^{+}_A (\V{k}'')
\left[ k_\perp'' \zhat \cdot (\kphat' \times \kphat'') \right]
\left[ \frac{k_\perp''}{k_\perp}+ \frac{k_\perp'}{k_\perp}(\kphat' \cdot \kphat'')
\right]
\label{eq:zpa_xy}
\end{equation}
\begin{equation}
\frac{\partial z^+_P (\V{k})}{\partial t} 
=- i\sum_{\V{k}=\V{k}'+\V{k}''}
 z^{-}_A  (\V{k}') z^{+}_P  (\V{k}'') \left[  k_\perp'' \zhat \cdot (\kphat' \times \kphat'') \right] 
  \left[  \frac{k_\perp k_\perp''}{k k''}\right] 
\label{eq:zpp_xy}
\end{equation}
The time evolution equations for $z^-_A(\V{k})$ and $z^-_P (\V{k})$
are the same as \eqsref{eq:zpa_xy}{eq:zpp_xy} with each
$z^+$ replaced by $z^-$, and vice versa.

The dramatically simplified \eqsref{eq:zpa_xy}{eq:zpp_xy} in
the perpendicular 2D limit imply two major changes in the physical
behavior compared to the full 3D case: (i) the linear term disappears,
eliminating the propagation of \Alfven and pseudo-\Alfven waves along
the (out-of-plane) equilibrium magnetic field, and implying that the
turbulence is always strong; and (ii) the nonlinear energy transfer
becomes entirely controlled by the \Alfvenic fluctuations, with the
pseudo-\Alfvenic fluctuations relegated to a passive role. Below we
discuss the implications of these limitations on the physical
behavior of incompressible MHD turbulence in the perpendicular 2D
limit.

In an incompressible MHD plasma, magnetic tension is the restoring
force that couples the magnetic field and velocity fluctuations and
leads to wave propagation of the Elsasser fields, $\V{z}^\pm$, along
the equilibrium magnetic field.  Without the possibility of any variation
along the equilibrium magnetic field (when $k_\parallel =0$, as implied by
the perpendicular 2D limit), there is no restoring force, and
therefore no wave behavior, only fluctuations within the perpendicular
2D plane. Note that the response governed by the linear term is the
lowest-order, fundamental response of the plasma to any perturbation
that varies along the field, and it is entirely absent in the
perpendicular 2D limit; in fact, there is no dynamical evolution at
all without the nonlinear term.  In this limit, the Elsasser fields
$\V{z}^\pm$ lose the physical meaning that they represent
finite-amplitude waveforms that travel up or down the equilibrium magnetic
field.

The elimination of wave physics in the perpendicular 2D limit has the
significant implication that it is not possible to have a state of
\emph{weak} incompressible MHD turbulence
\citep{Sridhar:1994}, and that the resulting
turbulence is always \emph{strong} incompressible MHD turbulence
\citep{Goldreich:1995} for any magnitude of fluctuations\footnote{Note that
the characteristic timescale for the evolution of the turbulence
increases with decreasing turbulent fluctuation amplitude.  But, since
no other timescale exists in perpendicular 2D limit of incompressible
MHD turbulence, the turbulent evolution will be equivalently strong over
similar periods of evolution when normalized by that characteristic
timescale.}  of the Elsasser fields $\V{z}^\pm$. The strength of the
turbulence can be measured by the nonlinearity parameter $\chi$
\citep{Goldreich:1995,Howes:2011b}, defined as the ratio of the
magnitude of the nonlinear to the linear term in
\eqref{eq:elsasserpm},
\begin{equation}
\chi \equiv \frac{\left|\V{z}^{\mp}\cdot \nabla \V{z}^{\pm}\right|}
{\left| \V{v}_A \cdot \nabla \V{z}^{\pm} \right|}.
\end{equation}
A state of weak turbulence occurs in the limit of weak nonlinearity,
$\chi \ll 1$, and a state of strong turbulence occurs for $\chi
\gtrsim 1$. In the perpendicular 2D limit, the denominator of
$\chi$ is always zero, so the turbulence is always strong. In other
words, the nonlinear terms always dominate the lowest order evolution
of the turbulent plasma dynamics. This characteristic of
incompressible MHD turbulence in the perpendicular 2D limit is similar
to the case of hydrodynamic turbulence, as  discussed in 
\secref{sec:perp2dhd} below.

Recalling that the independent conservation of $|\V{z}^+|^2$ and
$|\V{z}^-|^2$ implies that the Elsasser fields exchange no energy with
each other, \eqref{eq:zpa_xy} indicates that the nonlinear
interactions lead to energy transfer to $z_A^+(\V{k})$ only from
$z_A^{+}(\V{k}'')$, and \eqref{eq:zpp_xy} yields nonlinear energy
transfer to $z_P^+(\V{k})$ only from $z_P^{+}(\V{k}'')$.
Significantly, the energy transfer among these downward waves in this
perpendicular 2D limit is mediated \emph{only} by \Alfvenic
fluctuations $z_A^{-}(\V{k}')$, and these \Alfvenic fluctuations lose
no energy in the process.  Therefore, the pseudo-\Alfvenic
fluctuations exert \emph{no} influence on the nonlinear evolution of
the turbulence, but instead are passively advected by the \Alfvenic
fluctuations in the turbulence.  The \Alfvenic fluctuations, on the
other hand, are unaffected by the presence or absence of
pseudo-\Alfvenic fluctuations.

\subsubsection{Relation to 2D Hydrodynamics}
\label{sec:perp2dhd}
One may perform an identical projection of the Euler equations for
incompressible hydrodynamics,
\begin{equation}
\frac{\partial \V{v}}{\partial t} 
=-  \V{v}\cdot \nabla \V{v} -\nabla P/\rho_0,
\label{eq:euler}
\end{equation}
\begin{equation}
\nabla\cdot  \V{v}=0
\label{eq:divhd}
\end{equation} 
to obtain a comparable set of equations. In this case, we again use
the direction $\zhat$ to define the two nonzero polarizations of the
velocity field, $\eone= \zhat\times \V{k}$ and $\etwo=\hat{\V{k}}
\times (\zhat
\times \hat{\V{k}})$.  The lack of any preferred direction in
hydrodynamics implies isotropy, so the choice of the $\zhat$ direction
here is arbitrary, but is chosen to facilitate direct comparison of
the resulting equations to the incompressible MHD case.

We follow an analogous procedure to that presented in
\secref{sec:projection} to obtain two scalar equations for the evolution of
the two nonzero components of a single Fourier plane-wave
mode. Simplifying to the perpendicular 2D limit that allows variation
only in the $(x,y)$ plane, these equations reduce to
\begin{equation}
\frac{\partial v_1 (\V{k})}{\partial t} 
=- i\sum_{\V{k}=\V{k}'+\V{k}''}
v_1(\V{k}')v_1 (\V{k}'') \left[  k_\perp'' \zhat \cdot (\kphat' \times \kphat'') \right] 
\left[ \frac{k_\perp''}{k_\perp}+ \frac{k_\perp'}{k_\perp}(\kphat' \cdot \kphat'')
\right]
\label{eq:hd1_xy}
\end{equation}
\begin{equation}
\frac{\partial v_2 (\V{k})}{\partial t} 
=- i\sum_{\V{k}=\V{k}'+\V{k}''}
 v_1  (\V{k}') v_2 (\V{k}'') \left[  k_\perp'' \zhat \cdot (\kphat' \times \kphat'') \right]
 \left[  \frac{k_\perp k_\perp''}{k k''} \right]  
\label{eq:hd2_xy}
\end{equation}
where the Fourier coefficient for an arbitrary velocity fluctuation is
given by $\V{v}(\V{k}) = v_1 (\V{k})\eone +v_2 (\V{k})\etwo$ when
projected on the orthonormal basis $(\hat{\V{k}},\eone,\etwo)$, and
the wavevector $\V{k}=\V{k}_\perp=k_x \xhat+k_y \yhat$. In this 2D
limit, the polarization vectors reduce to $\eone =\zhat \times
\hat{\V{k}}_\perp$ and $\etwo =\zhat$, analogous to the polarization directions
for \Alfvenic and pseudo-\Alfvenic fluctuations in the perpendicular
2D limit of incompressible MHD.

The very significant similarities between 2D hydrodynamics and the
perpendicular 2D limit of incompressible MHD are clear by comparing
\eqref{eq:hd1_xy} to \eqref{eq:zpa_xy} and \eqref{eq:hd2_xy} to
\eqref{eq:zpp_xy}. With neither system containing a linear term, no wave 
propagation occurs in either system, and the turbulence in both of the
systems is always strong, independent of the amplitude of the
turbulent fluctuations when the timescales are appropriately
normalized.  In addition, the nonlinear evolution is controlled by the
\Alfvenic fluctuations in the perpendicular 2D limit of incompressible
MHD and by the $v_1$ polarized fluctuations in the 2D hydrodynamic
limit. Both of these modes have their polarization vectors contained
within the plane represented by the 2D spatial domain, $\ea =\zhat
\times \hat{\V{k}}_\perp$ and $\eone =\zhat \times
\hat{\V{k}}_\perp$.  The one  distinction between the  
perpendicular 2D incompressible MHD and 2D hydrodynamic systems is
that, in the incompressible MHD system, the $\V{z}^+$ fluctuations
cascade the $\V{z}^-$ fluctuations, and vice versa, while, in the
hydrodynamic system, the turbulent velocity $\V{v}$ cascades itself.

\subsubsection{Relation to Reduced MHD}
\label{sec:perp2drmhd}
The development of anisotropy is a widely recognized property of
magnetized plasma turbulence, supported by laboratory experiments
\citep{Robinson:1971,Zweben:1979,Montgomery:1981}, numerical
simulations
\citep{Shebalin:1983,Cho:2000,Maron:2001,Cho:2004,Cho:2009,TenBarge:2012a},
and solar wind observations
\citep{Belcher:1971,Sahraoui:2010b,Narita:2011,Roberts:2013}.  Even
for turbulence driven isotropically ($k_\perp
\sim k_\parallel$) at a large scale $L$, at perpendicular scales 
sufficiently smaller than the driving scale, $k_\perp L \gg 1$, the
inherently anisotropic energy transfer in plasma turbulence leads to
small-scale turbulent fluctuations that are highly elongated along the
direction of the magnetic field, described by the \emph{anisotropic
limit} $k_\perp\gg k_\parallel$. In this anisotropic limit, one may
derive a reduced set of equations for compressible MHD called
\emph{reduced MHD} \citep{Strauss:1976,Montgomery:1981,Montgomery:1982,Schekochihin:2009}.

As shown in \citet{Howes:2013a}, the equations for the evolution of
the \Alfvenic fluctuations in incompressible MHD---given by
\eqref{eq:zpa_final} and the complementary equation for $z^-_A
(\V{k})$---become identical to the equations of reduced MHD when the
anisotropic limit, $k_\perp\gg k_\parallel$, is adopted. This can be
clearly illustrated by assuming an ordering $k_\parallel/k_\perp \sim
\epsilon \ll 1$ and simplifying \eqref{eq:zpa_final}. Comparing the
magnitude of the four nonlinear terms on the right-hand side of
\eqref{eq:zpa_final}, the second and third terms are smaller than the
first by a factor of $\epsilon$, and the fourth term is smaller by a
factor of $\epsilon^2$. The resulting lowest-order equation is
\begin{equation}
\frac{\partial z^+_A (\V{k})}{\partial t} 
- i k_\parallel v_A z^+_A (\V{k}) =- i\sum_{\V{k}=\V{k}'+\V{k}''}
 z^{-}_A(\V{k}') z^{+}_A(\V{k}'') \left[ k_\perp'' \zhat
\cdot (\kphat' \times \kphat'') \right]
\left[ \frac{k_\perp''}{k_\perp}+ \frac{k_\perp'}{k_\perp}(\kphat' \cdot \kphat'')
\right]
\label{eq:rmhd}
\end{equation}
This equation is the same as the evolution equation for the \Alfvenic
fluctuations in the perpendicular 2D limit given by \eqref{eq:zpa_xy},
with one significant exception:  the  linear term is retained in the
anisotropic limit of reduced MHD!

The linear term is retained in the equations for reduced MHD because,
in many magnetized plasma systems of interest, the magnetic field
fluctuations are small compared to the equilibrium magnetic field,
corresponding to an ordering $|\V{z}^\pm|/v_A \sim \epsilon \ll
1$. Consequently, the magnitude of the linear term is approximately
the same order as the magnitude of the dominant nonlinear term,
$\left| \V{v}_A \cdot
\nabla \V{z}^{\pm} \right| \sim \left|\V{z}^{\mp}\cdot \nabla
\V{z}^{\pm}\right|$, a condition known as \emph{critical balance} in
the modern theory for anisotropic plasma turbulence
\citep{Higdon:1984a,Goldreich:1995,Boldyrev:2006}. In terms of 
\eqref{eq:rmhd} above, the ratio of the remaining nonlinear term to the linear
term (the nonlinearity parameter) is $\chi \sim k_\perp'' z^{-'}/
(k_\parallel v_A ) \sim 1$. Both the linear and nonlinear terms make
important contributions to the turbulent evolution in this limit.

In summary, the reduced MHD equations for the evolution of the
\Alfvenic component of compressible MHD turbulence in the anisotropic
limit $k_\perp \gg k_\parallel$, given by \eqref{eq:rmhd}, is very
similar to the perpendicular 2D limit of incompressible MHD given by
\eqref{eq:zpa_xy}. The significant difference is the presence of the
linear term in \eqref{eq:rmhd} for reduced MHD, and this difference
has important implications for the modeling of the turbulent plasma
dynamics. In reduced MHD, the turbulence consists of finite-amplitude
\Alfven waves propagating up and down the equilibrium magnetic field,
where each wave interacts nonlinearly with the counterpropagating
waves
\citep{Iroshnikov:1963,Kraichnan:1965,Howes:2012b,Howes:2013a,Nielson:2013a,
  Howes:2013b,Drake:2013}. These nonlinear interactions drive an
anisotropic turbulent cascade, with energy preferentially transferred
to small perpendicular scales \citep{Goldreich:1995,Boldyrev:2006}.
The existence of an anisotropic cascade in magnetized plasma
turbulence is well supported by laboratory experiments
\citep{Robinson:1971,Zweben:1979,Montgomery:1981}, numerical
simulations
\citep{Shebalin:1983,Cho:2000,Maron:2001,Cho:2004,Cho:2009,TenBarge:2012a},
and solar wind observations
\citep{Belcher:1971,Sahraoui:2010b,Narita:2011,Roberts:2013}.  It is
proposed that a state of critical balance between the linear and
nonlinear terms determines the properties of this anisotropic cascade
\citep{Goldreich:1995,Boldyrev:2006}. Therefore, the dynamics of
plasma turbulence inherently involves three-dimensional physics,
specifically the dominant nonlinearity that requires both
perpendicular dimensions and the wave physics that requires the
parallel dimension.

It is worthwhile pointing out that it is sometimes stated that reduced
MHD is not fully three-dimensional---this is incorrect. Indeed,
reduced MHD describes the fully three-dimensional dynamics of
anisotropic fluctuations in a magnetized plasma.  The perpendicular 2D
limit of incompressible MHD, on the other hand, models turbulence consisting
not of counterpropagating waves, but of non-propagating fluctuations
that have either \Alfvenic or pseudo-\Alfvenic polarizations. The
linear physics that plays a role in defining the anisotropic nature of
the cascade is absent in this perpendicular 2D limit, eliminating the
important wave-like properties of the turbulent fluctuations.

\subsection{In-Plane 2D Limit}
\label{sec:inplane2d}
In this limit, spatial variations may occur only in the $(x,z)$ plane,
so the possible wavevectors are $\V{k}=k_x \xhat + k_\parallel \zhat$,
where the equilibrium magnetic field is $\V{B}_0=B_0 \zhat$.  As a
result, the \Alfven wave polarization simplifies to $\ea =\yhat$ and
the pseudo-\Alfven wave polarization simplifies to $\ep =\khat
\times \yhat$. Since only one dimension exists in the perpendicular plane, 
$\hat{\V{k}}_\perp=\xhat$, it implies that the cross product between
the perpendicular components of any two wavevectors is always zero,
$\kphat' \times \kphat''=0$. Consequently,
\eqsref{eq:zpa_final}{eq:zpp_final} simplify   significantly  to yield
\begin{equation}
\frac{\partial z^+_A  (\V{k})}{\partial t} 
- i k_\parallel {v}_A z^+_A  (\V{k})
=- i\sum_{\V{k}=\V{k}'+\V{k}''}
 z^{-}_P  (\V{k}') z^{+}_A  (\V{k}'')
\left[   k_\parallel''\frac{k_\perp'}{k'}  
- k_\perp'' \frac{k_\parallel'}{k'}(\kphat' \cdot \kphat'') \right] 
 \left[ \frac{k_\perp''}{k_\perp}+ \frac{k_\perp'}{k_\perp}(\kphat' \cdot \kphat'') \right]
\label{eq:zpa_xz}
\end{equation}
\begin{eqnarray}
\lefteqn{\frac{\partial z^+_P  (\V{k})}{\partial t} 
- i k_\parallel {v}_A z^+_P  (\V{k})
=} \label{eq:zpp_xz}
\\
&&- i\sum_{\V{k}=\V{k}'+\V{k}''} 
  z^{-}_P  (\V{k}')z^{+}_P (\V{k}'') \left[   k_\parallel''\frac{k_\perp'}{k'}  
- k_\perp'' \frac{k_\parallel'}{k'}(\kphat' \cdot \kphat'') \right] 
\left[ \frac{k_\perp k_\perp''}{k k''} + 
 \frac{k_\parallel k_\parallel'' k_\perp''}{k k'' k_\perp}
+  \frac{k_\parallel k_\parallel''k_\perp'}{k k''k_\perp}( \kphat' \cdot \kphat'') \right]
\nonumber
\end{eqnarray}
The major change in the physical behavior of the turbulent plasma in
the in-plane 2D limit, demonstrated by \eqsref{eq:zpa_xz}{eq:zpp_xz},
is that the nonlinear energy transfer becomes entirely controlled by
the pseudo-\Alfven waves, with the \Alfven waves relegated to a
passive role.

The nonlinear dynamics of the in-plane 2D limit is found to be the
complement of the perpendicular 2D limit examined in
\secref{sec:perp2d}.  Since the oppositely propagating Elsasser fields exchange no 
energy with each other, \eqref{eq:zpa_xz} indicates that the nonlinear
interactions lead to energy transfer to $z_A^+(\V{k})$ only from
$z_A^{+}(\V{k}'')$, and \eqref{eq:zpp_xz} yields nonlinear energy
transfer to $z_P^+(\V{k})$ only from $z_P^{+}(\V{k}'')$.  The energy
transfer in the in-plane 2D limit is mediated \emph{only} by the
pseudo-\Alfven waves $z_P^{-}(\V{k}')$;  the \Alfven waves
exert \emph{no} influence on the nonlinear evolution of the
turbulence, but instead are passively advected by the pseudo-\Alfven
waves in the turbulence.

It has been shown that in-plane 2D simulations of incompressible MHD
turbulence develop an anisotropic cascade with $k_\perp \gg
k_\parallel$ \citep{Shebalin:1983}. In this anisotropic limit, the
nonlinearity parameter arising from the ratio of the remaining
nonlinear term to the linear term for both
\eqsref{eq:zpa_xz}{eq:zpp_xz} has an order of magnitude of $\chi \sim
k_\parallel z^-_P / (k_\parallel v_A) \sim z^-_P / v_A$.  Therefore,
strong turbulence in this limit demands $ z^-_P / v_A \sim 1$,
requiring significantly larger amplitudes to obtain strong turbulence
than the case of anisotropic \Alfvenic turbulence in the reduced MHD,
which demands only $ z^-_A / v_A \sim k_\parallel /k_\perp \ll 1 $.

\subsection{Oblique 2D Limit}
\label{sec:oblique2d}
The oblique 2D limit is significantly more complicated to analyze than
the either the perpendicular 2D or in-plane 2D limits, so a full
treatment of this limit is left for a more complete, subsequent
investigation. Here we simply point out an azimuthal asymmetry that
cannot be avoided in the oblique 2D limit.

In this limit, we allow spatial variations to occur only in the
$(x,y)$ plane, so the possible wavevectors are $\V{k}=k_x \xhat + k_y
\yhat$, but the equilibrium magnetic field is inclined by an angle
$\theta$ toward the $\xhat$ direction away from the normal to the
$(x,y)$ plane, $\V{B}_0=B_0 \sin \theta \xhat + B_0 \cos \theta
\zhat$. The inclination angle is typically taken to be small,
$\theta\ll 1$, although this not strictly necessary.  In this case,
the parallel component of the wavevector is given by $k_\parallel =
\V{k} \cdot \V{B}_0/B_0 = k_x \sin \theta$ and the perpendicular
component is given by $\V{k}_\perp = k_x (1-\sin
\theta) \xhat + k_y \yhat$. 

In magnetized plasma turbulence, the magnetic field establishes a
preferred direction, but in the plane perpendicular to the magnetic
field, the statistical distribution of turbulent power is believed to
be axisymmetric about the magnetic field direction. Thus, there should
arise no azimuthal dependence in the turbulent
fluctuations. Specifically, if one chooses a set of wavevectors with a
constant magnitude of both the parallel and perpendicular components,
but spanning the full range of azimuthal angles about the field, one
should observe no statistical variation.  

Let us consider the closest equivalent to this situation that can be
represented in the oblique 2D limit, examining the set of wavevectors
that constitute a ring of constant radius $k_0$ in the $(x,y)$ plane,
$\V{k}(\phi)=k_0 \cos \phi \xhat + k_0 \sin \phi \yhat$. In this case,
the parallel and perpendicular components of the wavevector are
$k_\parallel = k_0 \sin \theta \cos \phi$ and $\V{k}_\perp = k_0
(1-\sin \theta) \cos \phi \xhat + k_0 \sin \phi \yhat$.  It is clear
that the parallel component of the wavevector contains an unavoidable
sinusoidal dependence on the azimuthal angle $\phi$, establishing a
connection between the direction of the perpendicular component of the
wavevector and the value of the parallel component.  For example, the
plane-wave mode that has a perpendicular wavevector component in the
$\xhat$ direction ($\phi=0$) describes an \Alfven or pseudo-\Alfven
wave that propagates up the magnetic field, while the mode in which
the perpendicular component points in the $-\xhat$ direction
($\phi=\pi$) describes a wave that propagates down the magnetic field.
If the plane-wave mode has a perpendicular component in the $\yhat$
($\phi=\pi/2$) or $-\yhat$ ($\phi=3\pi/2$) direction, the parallel
component of the wavevector is zero and the wave does not propagate at
all. The relation between the direction of the perpendicular component
of the wavevector and the value of the parallel component---a
connection arising from the limitation of the range of possible
wavevectors to the oblique 2D plane---will almost certainly introduce
artificial azimuthal asymmetries in the resulting turbulent dynamics
that will be difficult to interpret.

\section{Discussion}
\label{sec:discuss}
It is often asserted or implicitly assumed, without justification,
that the results of two-dimensional investigations of plasma
turbulence are applicable to the real, three-dimensional plasma
environments of interest. It is plausible that this belief is based on
the observed fact that, with a sufficient amplitude of large-scale
fluctuations (whether driven or decaying), it is indeed possible to
generate a turbulent cascade in any of the 2D limits discussed in
\secref{sec:2dlimits}.  The analysis of ideal incompressible MHD turbulence 
presented here demonstrates unequivocally that the character of the
nonlinearities underlying the turbulent cascade is dramatically
limited under reduced dimensionality. In both the perpendicular 2D and
in-plane 2D limits discussed here, the reduction in dimensionality
eliminates all but one of the channels of nonlinear energy transfer
that are possible in a three-dimensional treatment.

That two-dimensional treatments of ideal incompressible MHD turbulence
should demonstrate significantly different physical behavior from the
three-dimensional case should not be particularly surprising.  In
hydrodynamic turbulence, the cascade properties of the inviscid
invariants differ dramatically between the 3D and 2D cases: in 3D
hydrodynamic turbulence, there arises a direct cascade to small scales
of energy and kinetic helicity; in 2D hydrodynamic turbulence, the
energy cascades inversely to large scales while the enstrophy
undergoes a direct cascade
\citep{Kraichnan:1967,Batchelor:1969,Kraichnan:1980}.
Yet the three ideal invariants of incompressible MHD---the energy,
cross helicity, and magnetic helicity
\citep{Woltjer:1958a,Woltjer:1958b,Matthaeus:1982b}---exhibit similar cascade  
directions in both the 2D and 3D cases, with a direct cascade of
energy and cross helicity and an inverse cascade of magnetic helicity
(or the anastrophy in 2D) \citep{Biskamp:2003}.  It is possible that
this similarity of cascade directions in MHD has also fueled the belief
that 2D MHD turbulence simulations can be safely used to model 3D
plasma environments.

In the perpendicular 2D limit of incompressible MHD discussed in
\secref{sec:perp2d}, the linear term and
three of the four possible nonlinearities are eliminated from the
general \eqsref{eq:zpa_final} {eq:zpp_final} that govern the evolution
of the \Alfvenic and pseudo-\Alfvenic fluctuations. The elimination of
wave physics by dropping the linear term has significant implications
for the modeling of plasma turbulence. First, the development of the
typical wavevector anisotropy $k_\perp \gg k_\parallel$ that is often
observed in magnetized plasma turbulence cannot be studied in the
perpendicular 2D limit since $k_\parallel=0 $ always. The concept of
critical balance in the modern theory for anisotropic MHD turbulence
is likewise prohibited since it depends on maintaining a balance
between the timescales associated with the linear and nonlinear terms.
Second, a state of weak MHD turbulence is not possible since the
nonlinear term always dominates the dynamics.  The perpendicular 2D
limit of incompressible MHD, in fact, has significant similarities
with the case of incompressible 2D hydrodynamic turbulence, as shown
in \secref{sec:perp2dhd}, for which a state of weak hydrodynamic
turbulence is precluded. In the perpendicular 2D limit, the nonlinear
energy transfer is controlled solely by the \Alfvenic fluctuations,
with no transfer of energy between the \Alfvenic and pseudo-\Alfvenic
components of the turbulence. The reduced MHD equations contain the
same dominant nonlinear terms as the perpendicular 2D limit but also
retain the important physical wave behavior associated with the linear
term. Therefore, the reduced MHD equations are preferable to the
perpendicular 2D limit of the incompressible MHD equations for the
study of MHD turbulence.

In the in-plane 2D limit of  incompressible MHD discussed in
\secref{sec:inplane2d}, the linear term is retained but, again, three 
of the four possible nonlinearities are eliminated from the general
\eqsref{eq:zpa_final} {eq:zpp_final}. In this case, the nonlinear energy transfer
is controlled only by the pseudo-\Alfven wave dynamics, with the
\Alfven waves relegated to a passive role.  Similar to the
perpendicular 2D limit, the \Alfven waves and pseudo-\Alfven waves
exchange no energy with each other. In the anisotropic limit $k_\perp
\gg k_\parallel$, the in-plane 2D limit requires significantly larger
turbulent wave amplitudes to realize a state of strong MHD turbulence. 

Recently, \citet{Tronko:2013} conducted an explicit comparison of the
properties of weak incompressible MHD turbulence for the in-plane 2D
and 3D cases. In contrast to the domination by local interactions and
the existence of a well-behaved Kolmogorov-Zakharov spectrum in 3D
incompressible MHD turbulence, for the in-plane 2D limit, the
interactions are nonlocal and a Kolmogorov-Zakharov spectrum is not
realizable. This study concludes with ``a warning that 2D and 3D MHD
turbulence are dramatically different, and one should be careful when
extrapolating the 2D results, e.g., numerical ones, to the 3D case.''

In the oblique 2D limit discussed in \secref{sec:oblique2d}, although
we have not presented a thorough analysis, we have demonstrated an
unavoidable azimuthal dependence of the wavevector components that
will almost certainly introduce artificial azimuthal asymmetries in
the resulting turbulent dynamics that will be difficult to interpret.
 
It is worthwhile noting the shared property of the perpendicular 2D
limit and in-plane 2D limit of incompressible MHD and of the 2D limit
of incompressible hydrodynamics that the components of the turbulent
fluctuations that are polarized in the out-of-plane direction do not
play a role in mediating the nonlinear energy transfer.  These
modes---the pseudo-\Alfvenic fluctuations in the perpendicular 2D
limit, the \Alfven waves in the in-plane 2D limit, and the $v_2$
component of 2D hydrodynamics---are passively cascaded by the
remaining nonlinearity in the equation of evolution.


All of these results lead to the conclusion that, although in any of
these 2D configurations at least one nonlinearity persists to drive a
turbulent cascade, the nature of the remaining nonlinearity is very
different depending on the 2D configuration that is chosen. \Alfvenic
fluctuations control the nonlinear physics of the turbulent cascade in
the perpendicular 2D limit, but pseudo-\Alfven waves control the
nonlinear physics of the turbulent cascade in the in-plane 2D limit.
The wave response governed by the linear term is entirely absent in
the perpendicular 2D limit.  Therefore, before the lessons learned
from 2D treatments of plasma turbulence can be applied to understand
the behavior in 3D plasma environments of interest, it is essential
that the applicability of the 2D approach is first established. The
equations of reduced MHD, which include the dominant perpendicular
nonlinearity, as in the perpendicular 2D limit, but also retain the
linear term governing the wave physics, are preferable compared to any
of the 2D approaches.

\section{Conclusion}
\label{sec:conc}
In this study, we have derived two scalar equations for the evolution of
the \Alfven wave and pseudo-\Alfven wave components of ideal
incompressible MHD plasma turbulence by projecting the equations onto
the polarization direction for each wave type. These equations
highlight the four possible nonlinearities responsible for the
turbulent cascade of energy and reveal the geometric properties that
lead to the elimination of many possible channels for nonlinear energy
transfer under reduced dimensionality. Inspection of the scalar
\eqsref{eq:zpa_final} {eq:zpp_final} demonstrates the inherently 
three-dimensional nature of incompressible MHD turbulence. 

We have explored the nonlinear properties of several two-dimensional
limits of the incompressible MHD equations. In the perpendicular 2D
limit, the wave physics that plays a crucial role in the development
of anisotropy is eliminated, and the nonlinear energy transfer is
controlled only by the \Alfvenic fluctuations. In the in-plane 2D
limit, the nonlinear energy transfer is controlled only by the
pseudo-\Alfven waves. In the oblique 2D limit, an unavoidable
azimuthal dependence will likely cause artificial azimuthal
asymmetries in the resulting turbulent dynamics. Therefore, none of
these 2D limits is sufficient to capture fully the rich
three-dimensional nonlinear dynamics critical to the evolution of
incompressible MHD turbulence.

Although it is true, in the anisotropic limit $k_\perp \gg
k_\parallel$ that naturally develops in magnetized plasma turbulence,
that the dominant nonlinearities are the same as those retained in the
perpendicular 2D limit, the dropping of the linear term in this limit
eliminates the physical behavior that is necessary to capture the
anisotropic cascade of energy.  In this sense,  a three-dimensional
treatment using the reduced MHD equations, which maintain the same
dominant nonlinearities but also retain the linear term, is preferable
to the perpendicular 2D limit.

Key properties of plasma turbulence that occur in the idealized
incompressible MHD description, including the development of an
anisotropic \Alfvenic turbulent cascade and the generation of current
sheets at small scales, persist under less restrictive plasma
conditions that require a more sophisticated kinetic description.
Therefore, the inherently three-dimensional nature of incompressible
MHD turbulence is expected to remain a general characteristic of the
kinetic turbulence that occurs under conditions relevant to
astrophysical plasma systems.


The work has been supported by NSF CAREER Award AGS-1054061, NSF
PHY-10033446, and NASA NNX10AC91G.


\end{document}